\documentclass[aps,twocolumn]{revtex4}
\usepackage{epsfig}

\newcommand{\be}{\begin{equation}}
\newcommand{\ee}{\end{equation}}
\newcommand{\ba}{\begin{array}}
\newcommand{\ea}{\end{array}}

\usepackage{epsfig}
\begin{document}

\title{Intramolecular phase separation of copolymer "bottle brushes":
No sharp phase transition but a tunable length scale}
\author{Hsiao-Ping Hsu, Wolfgang Paul, and Kurt Binder}
\affiliation{Institut f\"ur Physik, Johannes Gutenberg-Universit\"at Mainz\\
D-55099 Mainz, Staudinger Weg 7, Germany}

\date{\today}

\begin{abstract}
A lattice model for a symmetrical copolymer "bottle brush" molecule,
where two types (A,B) of flexible side chains are grafted with one
chain end to a rigid backbone, is studied by a variant of the
pruned-enriched Rosenbluth method (PERM), allowing for simultaneous
growth of all side chains in the Monte Carlo sampling.
Choosing repulsive binary interactions between unlike monomers
and varying the solvent quality, it is found that phase separation into
an $A$-rich part of the cylindrical molecule and a $B$-rich part can occur
only locally. Long range order (in the direction of the backbone)
does not occur, and hence the transition from the randomly mixed
state of the bottle brush to the phase separated structure is strongly
rounded, in contrast to corresponding mean field predictions. This lack of a phase
transition can be understood from an analogy with spin models in one space
dimension.
We predict that the range of microphase separation along the bottle brush
backbone can be controlled on the nanoscale by varying the solvent quality.
\end{abstract}

\maketitle

\section{Introduction and Motivation}

Macromolecules with a comb-like architecture, where side chains are 
grafted to a linear
(backbone) chain, find increasing interest: if the backbone is a rigid
polymer, its solubility and processibility is improved ~\cite{Ciferri,Wegner,Cao}; if the
grafting density of the side chains is very high, cylindrical 
"bottle brush"-shaped objects form, which under certain conditions show a 
thermally induced collapse transition to a spherical structure, providing 
interesting perspectives for the design of "molecular actuators"~\cite{Gunari};
if copolymer bottle brushes are created by dense grafting of two types of
side chains $(A,B)$ on the backbone, interesting "horseshoe" and
"meander"-like structures are observed, which is attributed to a local phase
separation along the backbone of the bottle brush~\cite{Stephan}.
Clearly, this creates additional possibilities when one considers the
desired~\cite{Stephan} use of such shape persistent cylindrical 
macromolecules as building blocks in molecular assemblies.
Thus, the problem of phase separation within a simple copolymer bottle
brush has also found attention both  by analytical theory~\cite{Stepanyan} and
computer simulation~\cite{deJong} , the possibility to create 
"Janus cylinders" (upper half of the cylinder containing the 
$B$ monomers, lower half containing the $A$ monomers) was
suggested, and predictions how the phase transition point from
the randomly mixed state (where both $A$ and $B$ monomers are
homogeneously distributed in the cylinder volume)
to the separated state depends on the chain length of the side chains $N$
were given~\cite{Stepanyan}.

\begin{figure}
\begin{center}
\epsfig{file=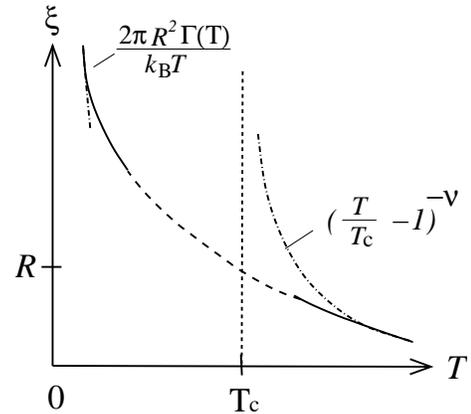, width=6.0cm, angle=0}
\caption{Crossover scaling behavior of the correlation length
$\xi$ as a function of temperature $T$. $\xi \propto (T/T_c-1)^{-\nu}$
for $T>T_c$, and $\xi \propto R^2 \Gamma(T)/ k_B T$ for
$T<T_c$.}
\label{cross}
\end{center}
\end{figure}

\begin{figure}
\begin{center}
\epsfig{file=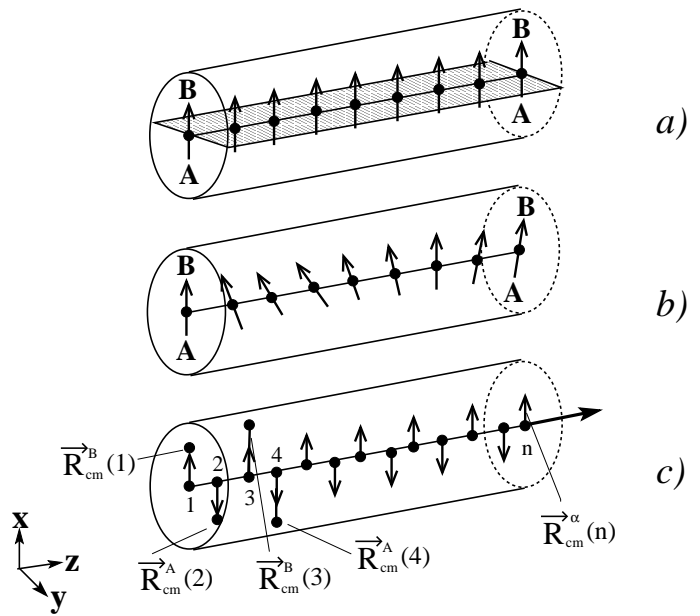, width=9.0cm, angle=0}
\caption{Perfect phase separation of side chains in a binary $(A,B)$
copolymer bottle brush with alternating grafting sequence $ABAB\cdots$
into a "Janus cylinder" structure implies formation of an interface
plane (part a), shaded) between the $A$-rich part (bottom) and the $B$-rich
part (top) of the cylindrical bottle brush. The local orientation
of this interface can be characterized by a vector oriented normal
to it (arrows). At nonzero but low temperature phase separation will
occur locally, but entropy will lead to long-wavelength fluctuations of
the orientation of this vector (b), destroying axial long range order 
along the direction of the backbone of the bottle brush. Since the
interface is not a sharp plane at nonzero temperature, but rather has
a finite width, a numerical characterization of the local orientation
of the interface is difficult. An essentially equivalent but numerically
unambiguous characterization of this "Janus cylinder"-type ordering is
achieved by calculating
the unit vector from the grafting site $(n)$ of each chain to the
projection of its center of mass position into the xy-plane
$\vec{R}_{cm}^\alpha(n)$, see part $(c)$.}
\label{cylinder}
\end{center}
\end{figure}

    In the case of flexible backbones there is a delicate
interplay between the local segregation between the $A$ and $B$
side chains and the global geometric structure (such as 
"horseshoe", etc.) of the bottle brush, requiring a detailed and 
accurate understanding of the phase separation in these molecules.
We reconsider this problem, performing Monte Carlo
simulations of a similar model as in Ref.~\cite{deJong}, but our analysis
focuses on a different aspect, that has been completely ignored
in the literature so far: single bottle brush molecules with
side chains of finite chain length $N$ and very long, ideally
infinitely long length of the backbone chain are quasi-one
dimensional (1d) objects. Statistical mechanics
of systems with short range
interactions implies quite generally~\cite{Domb,Baxter} that 
there cannot be any sharp phase transitions in $1d$ systems in
thermal equilibrium. A point in case is e.g. the $1d$ XY model,
a chain of spins where each spin $i$ is described by an angle $\varphi_i$
in the $xy$-plane, with $0 \leq \varphi_i \leq 2\pi$. 
Neighboring spins along the chain experience a coupling 
$- J \cos(\varphi_i-\varphi_j)$. While mean field theories predict that
ferromagnetic order along the chain occurs for temperatures
$T<T_{\rm c}^{\rm MF}=c J$, where $c$ is a constant of order 
unity~\cite{Domb}, an exact solution of this problem~\cite{Domb,Baxter}
shows that ferromagnetic long range order is unstable against 
long-wavelength fluctuations, and actually the ferromagnetic
correlation length $\xi$ grows completely gradually as the temperature
is lowered,
\be
     \xi= 2a (J/k_BT) \; ,   \label{xi}
\ee
$a$ being the lattice spacing. Thus $T_c=0$: note that
$\xi$ has to diverge when $T_c$ is approached, and the singularity of
$\xi$ predicted by mean field theory at $T=T_c^{MF}$ is completely
washed out. 
This consideration can be generalized to cylinders of cross
section area $\pi R^2$~\cite{Fisher}.
Then Eq.~(\ref{xi}) gets replaced by a more complicated behavior
that is sketched in Fig.~\ref{cross}. For $T>T_c$ (the critical
temperature of the corresponding bulk three-dimensional
system) $\xi \propto (T/T_c-1)^{-\nu}$, where $\nu$ is the appropriate
critical exponent, until $\xi$ is of the order of $R$.
Then the critical singularity is rounded off, and a crossover sets 
in to a relation analogous to Eq.~(\ref{xi})
\be
   \xi= 2\Gamma(T)\pi R^2/k_BT \propto (R^2/a)(J/k_BT)
\;, \enspace T \rightarrow 0\;. \label{xispin}
\ee
Note that the "spinwave stiffness" (helicity modules) 
$\Gamma(T)$ characterizes the cost of 
long wavelength order parameter rotations, and 
$\Gamma(T \rightarrow 0) \propto J$.
In the present work we shall provide
evidence that the correlation length describing
phase separation of a "Janus cylinder" type behaves qualitatively
similar to Eqs.~(\ref{xi}) and (\ref{xispin}), 
cf. fig.~\ref{cylinder}b. Then the Flory-Huggins parameter
$\chi_{AB}$ that describes the incompatibility between $A$ and $B$
monomers takes the role that $(J/k_BT)$ plays for the spin model.
However, the problem whether for $N \rightarrow \infty$ 
(and depending on the solvent condition) sharp phase transitions
are restored is nontrivial, and in particular, describing the extent of
rounding of the (mean field) transitions described in ~\cite{Stepanyan} 
by these long range fluctuations along the chain backbone remains 
a challenge.

\section{Model and simulation technique}

   As considered previously~\cite{Stepanyan,deJong}, the generic model for this problem 
treats a completely rigid rod, oriented along the $z$-axis of a simple 
cubic lattice, and the side chains of type $A$ and $B$ are grafted 
in an alternating way, every site of the backbone chain being the grafting 
site of a side chain end. 
Of course, creating a statistical
copolymer where the type of side chain along the backbone is
chosen at random would be more realistic, from the experimental
point of view. However, the quenched disorder thus introduced
would encumber a theoretical treatment considerably, therefore this has not been
considered previously~\cite{Stepanyan,deJong}, and hence 
we study here regularly 
alternating copolymers, for the sake of better comparison to
previous work. Also, we introduce a periodic boundary condition in
$z$-direction, to avoid end effects associated with a finite
backbone length.

   The side chains are self-and mutually avoiding random walks. 
Between monomers of the side chains and
monomers of the backbone chain only the excluded volume effect is considered.
We allow for nearest neighbor interactions $\epsilon_{AB}$,
$\epsilon_{AA}=\epsilon_{BB}=\epsilon$ between the respective pairs 
of monomers of the side chains, and hence the partition sum for this model is
\be
    Z=\sum q^{m_{AA}+m_{BB}} q_{AB}^{m_{AB}} \;, \label{partition}
\ee
where $q=\exp(-\epsilon/k_BT)$, $q_{AB}=\exp(-\epsilon_{AB}/k_BT)$,
and $m_{AA}$, $m_{BB}$, $m_{AB}$ are the numbers of non-bonded
occupied nearest neighbor monomer pairs $AA$, $BB$, and $AB$,
respectively. The sum in Eq.~(\ref{partition}) extends over all
possible configurations of these walks. Varying $q$ in the
range $1\leq q \leq 1.5$, we cover the full range from good
solvents ($\epsilon=0$) to poor solvents ($\epsilon=1.5$),
since the $\theta$-solvent corresponds to a choice~\cite{g97}
$q_\theta=\exp(-\epsilon/k_BT_\theta)\approx 1.3087$. We vary
$q_{AB}$ in the range $0 \leq q_{AB} \leq q$:
The case $q_{AB}=0$ corresponds to a very strong repulsion between
$A$ and $B$, while for $q_{AB}=q$ the chemical incompatibility 
vanishes $\left\{ \right .$recall that~\cite{Flory} $\chi_{AB} \propto
\epsilon_{AB}-(\epsilon_{AA}+\epsilon_{BB})/2$$\left . \right\}$.

\begin{figure*}
\begin{center}
$\begin{array}{c@{\hspace{0.5cm}}c}
\epsfig{file=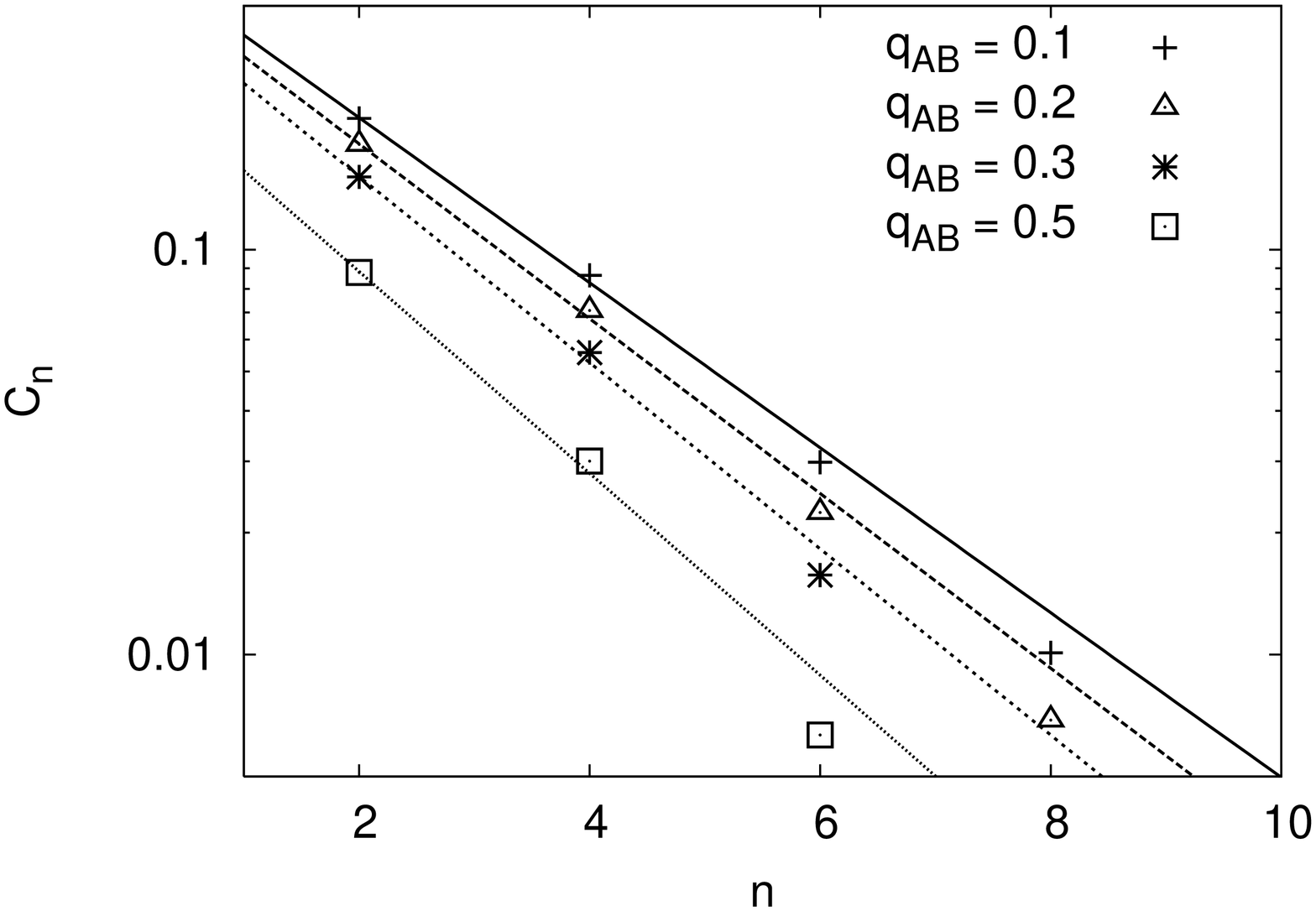, width=4.5cm, angle=270} &
\epsfig{file=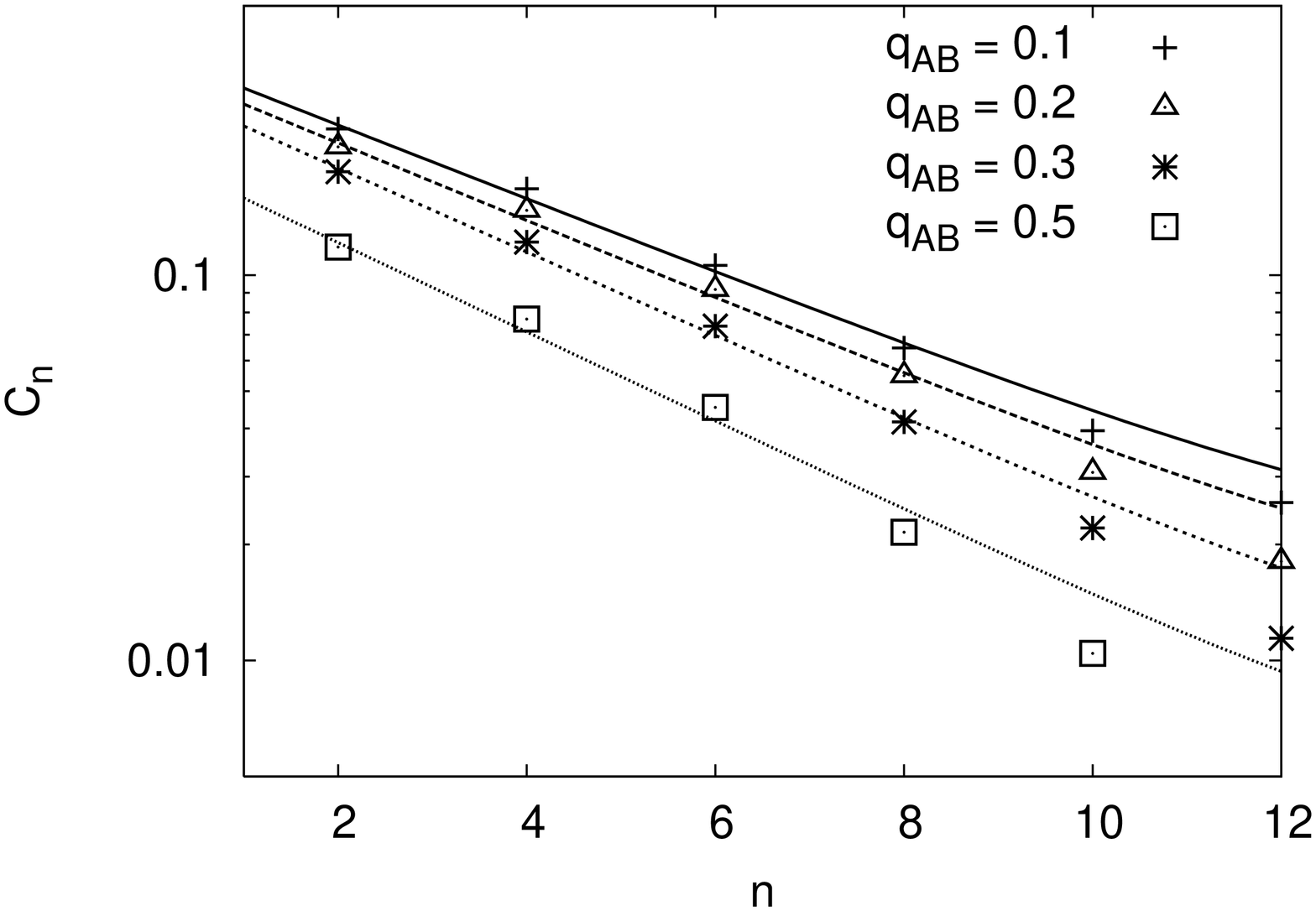, width=4.5cm, angle=270} \\
\end{array}$
\caption{Correlation function $C_n$ describing
the correlation of
local phase separation along the backbone plotted vs.\ the
distance $n$ on a semi-log scale, for good solvent conditions $(\epsilon=0)$,
$N=6$ (left part) and $N=18$ (right part). All data
are for $L_b=32$, and several choices of $q_{AB}$ are included,
as indicated. Note that
$C_n \equiv(<{\vec S}_i^A \cdot {\vec S}_{i+n}^A>+
<{\vec S}_{i+1}^B \cdot {\vec S}_{i+1+n}^B>)/2$.
Curves show the best fit to the data,
i.e. $C_n \sim [\exp(-n/\xi)+\exp(-(L_b-n)/\xi)]$}.
\label{cn-good}
\end{center}
\end{figure*}

\begin{figure*}
\begin{center}
$\begin{array}{c@{\hspace{0.5cm}}c}
\epsfig{file=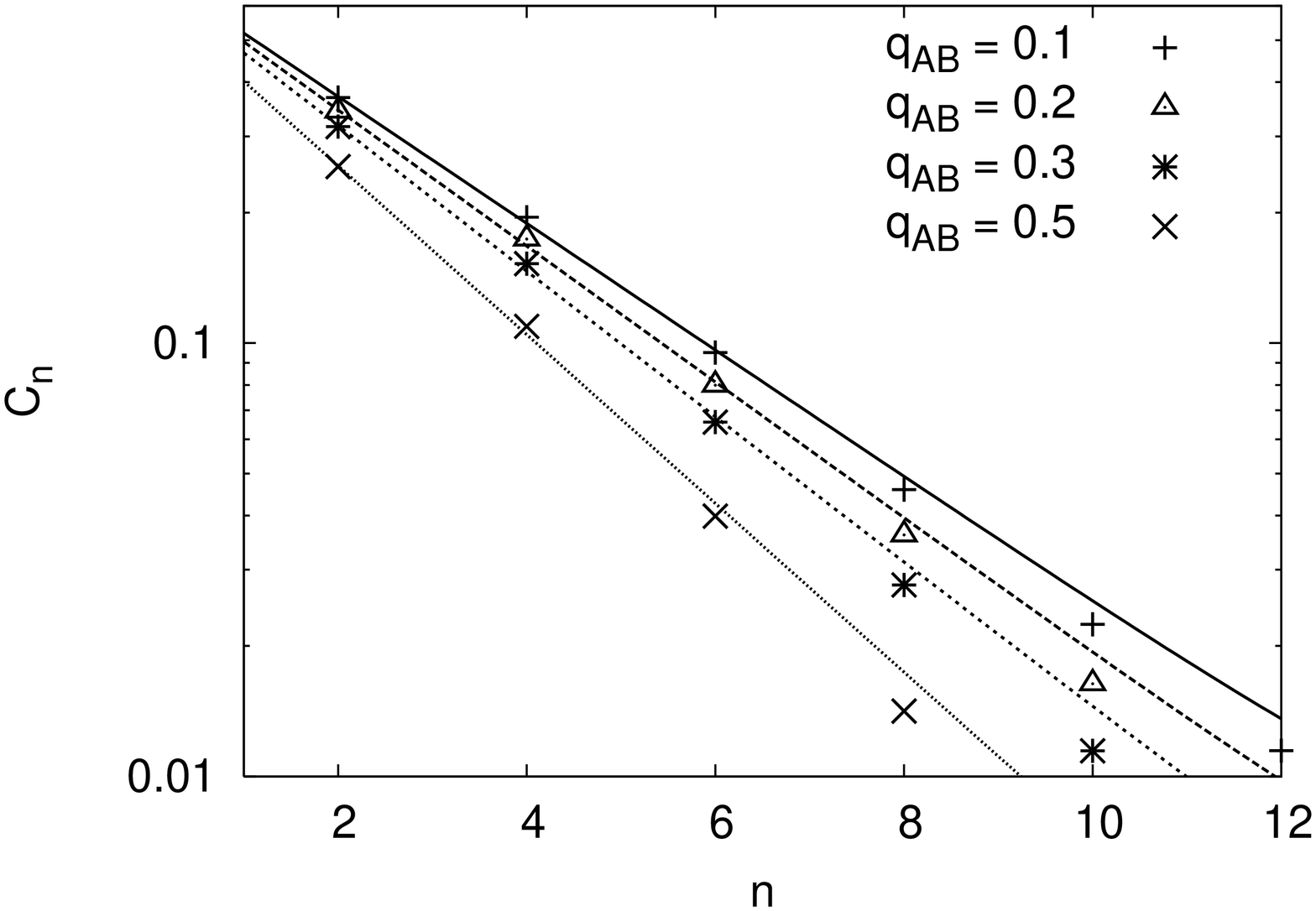, width=4.5cm, angle=270} &
\epsfig{file=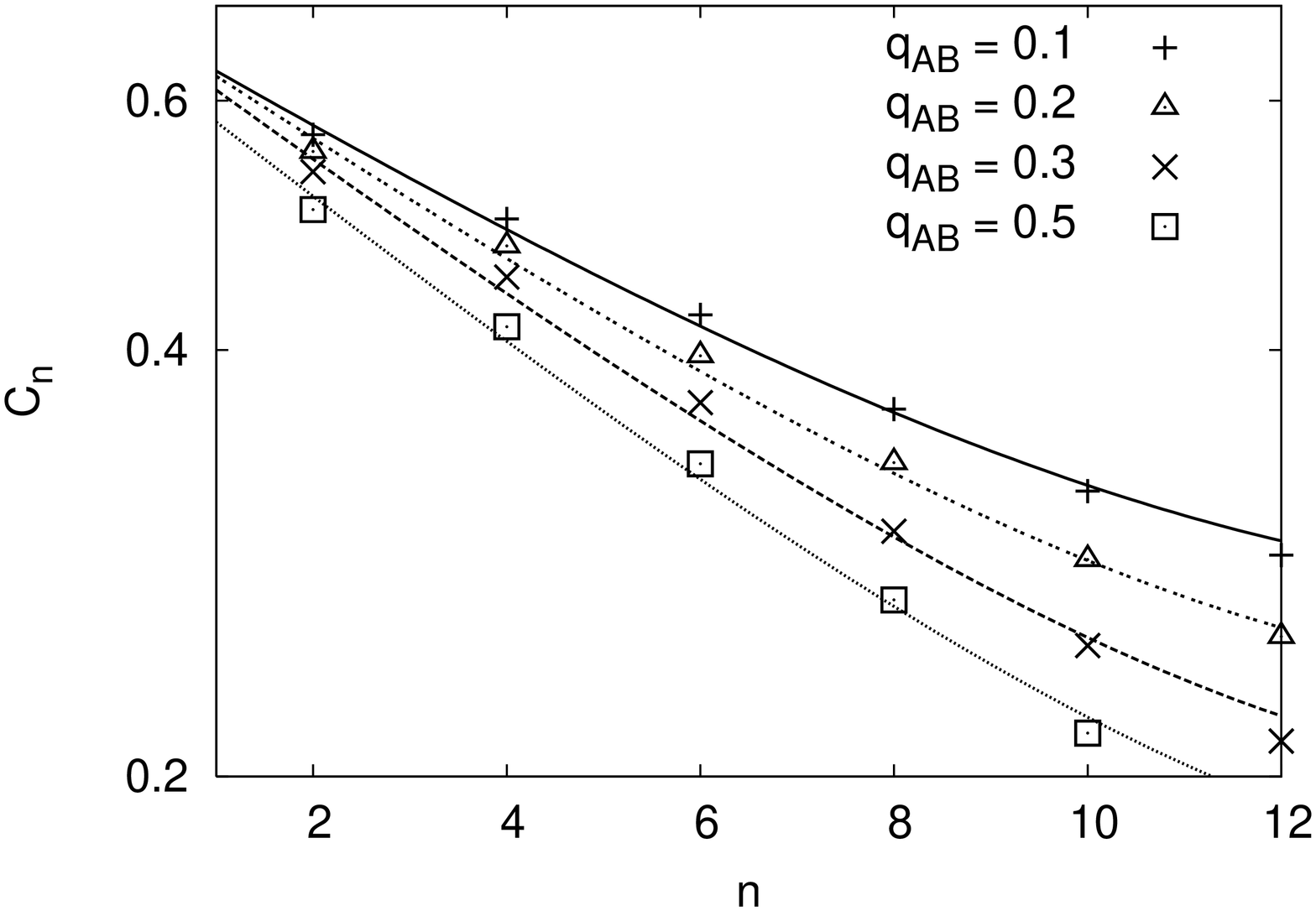, width=4.5cm, angle=270} \\
\end{array}$
\caption{Correlation function $C_n$ plotted vs.\ $n$ on a semi-log scale,
for a poor solvent ($q=1.5$), $L_b=32$, $N=6$ (left)
and $N=18$ (right). Several choices of $q_{AB}$ are included,
as indicated. Curves show the best fit to the data, 
i.e. $C_n \sim [\exp(-n/\xi)+\exp(-(L_b-n)/\xi)]$}.
\label{cn-poor}
\end{center}
\end{figure*}

   For our simulations, we use the pruned-enriched Rosenbluth model 
(PERM)~\cite{g97}. This is a biased chain growth algorithm with population
control. PERM has been applied very successfully both to linear
and branched polymers (stars~\cite{h04star1,h04star2}, lattice animals~\cite{h05anim}).
Similar to~\cite{h04star1,h04star2}, the bottle brush is generated by adding one 
monomer to each side chain until all side chains have the same number
of monomers, thus growing all side chains simultaneously. For efficiency,
side chains are grown with higher probabilities in the directions perpendicular to
the backbone and in the direction where there are more free next 
neighbor sites. This additional bias must be taken into account by
suitable weight factors. High statistics 
($10^5 \sim 10^6$ independent configurations)
is obtained by
using the population control (pruning/cloning)~\cite{g97,h04star1,h04star2,h05anim}. The backbone
length also is varied ($L_b=32$, $48$, $64$) to check
that finite size effects caused by too small $L_b$ are still negligible.

\begin{figure*}
\begin{center}
$\begin{array}{c@{\hspace{0.5cm}}c}
q_{AB}=1.5 & q_{AB}=0.1 \\
\epsfig{file=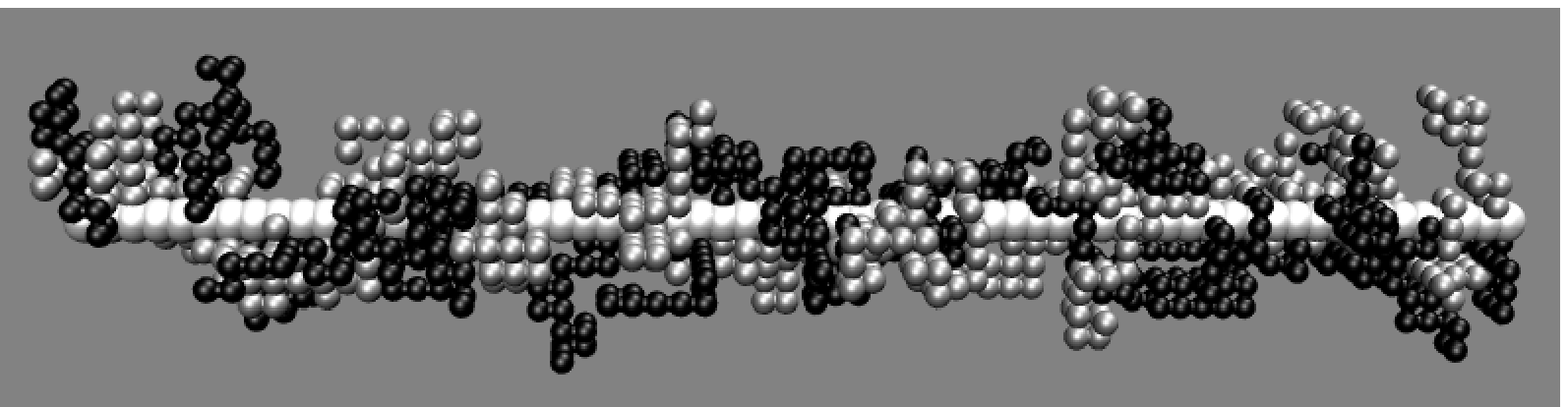, width=6.5cm, angle=0} & 
\epsfig{file=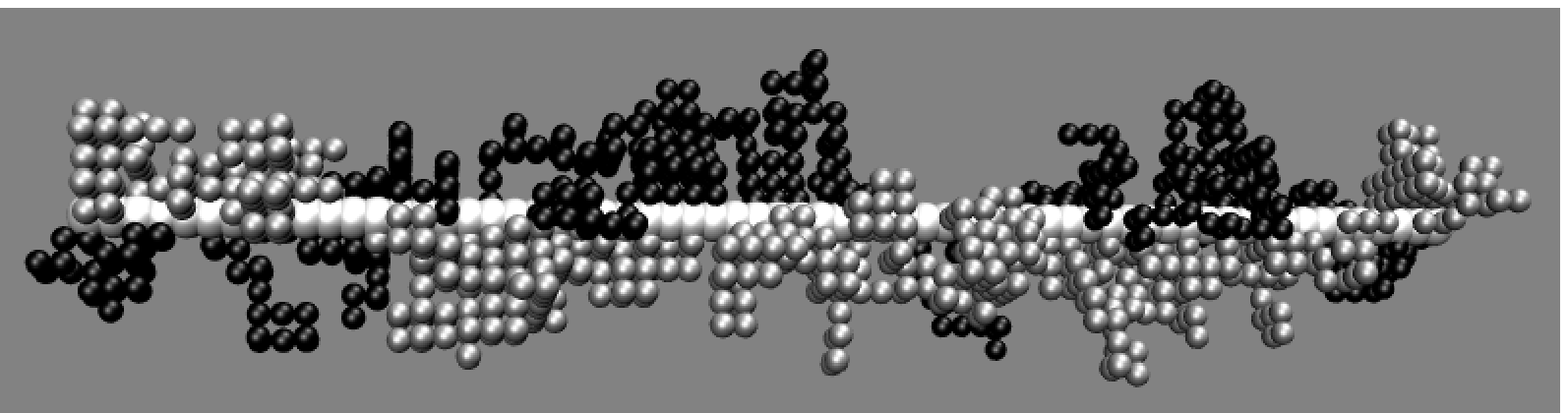, width=6.5cm, angle=0} \\
\end{array}$
\caption{Snapshots of the copolymer bottle brushes for a poor solvent ($q=1.5$),
$L_b=64$, $N=18$, $q_{AB}=1.5$ (mixed state) and $q_{AB}=0.1$
(state with local phase separation). 
Monomers $A$ , monomers $B$, and monomers on the backbone
are shown in black, gray, and white colors, respectively.}
\label{snapshot}
\end{center}
\end{figure*}

\begin{figure*}
\begin{center}
$\begin{array}{c@{\hspace{0.5cm}}c}
\epsfig{file=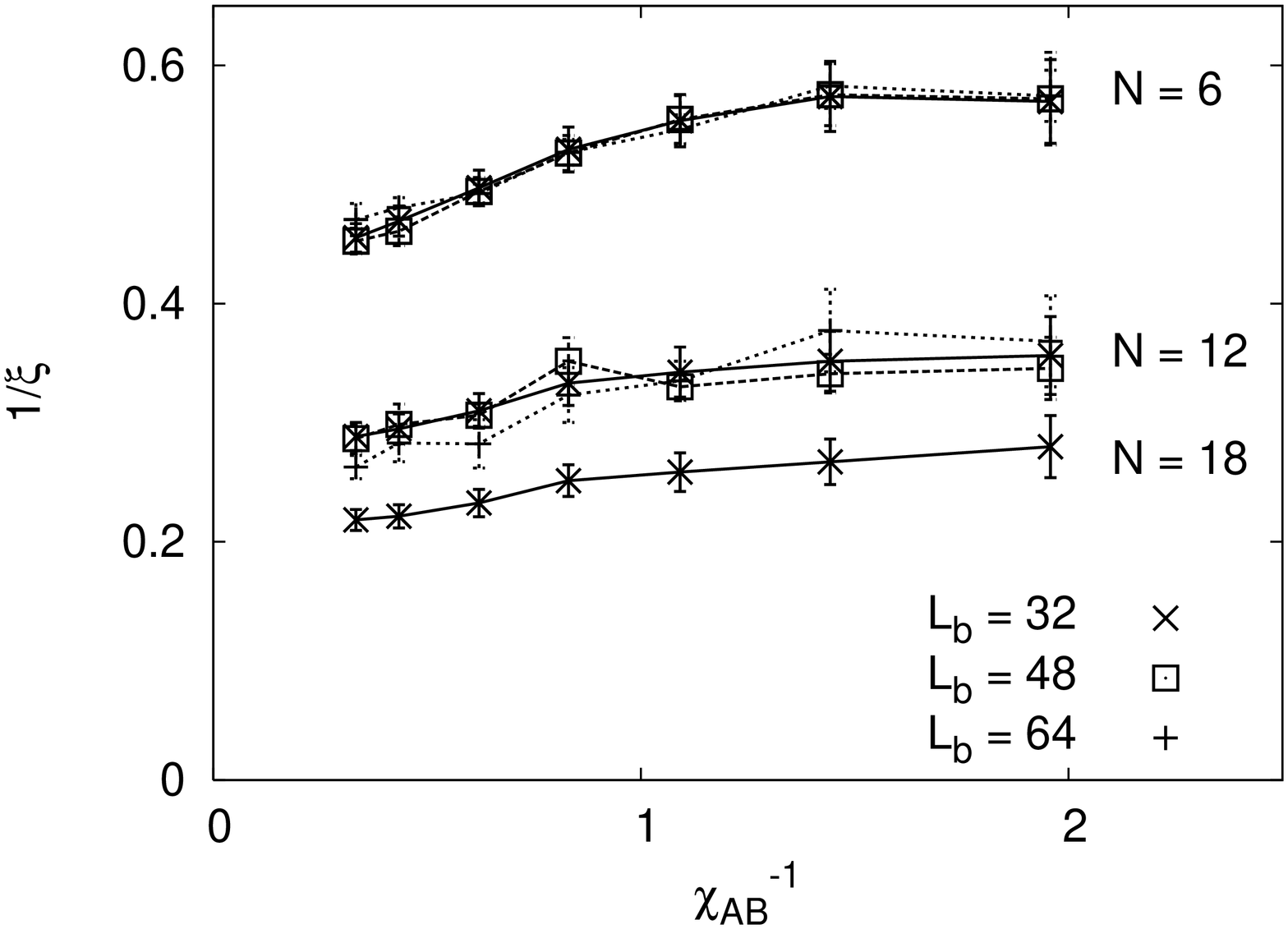, width=4.5cm, angle=270} &
\epsfig{file=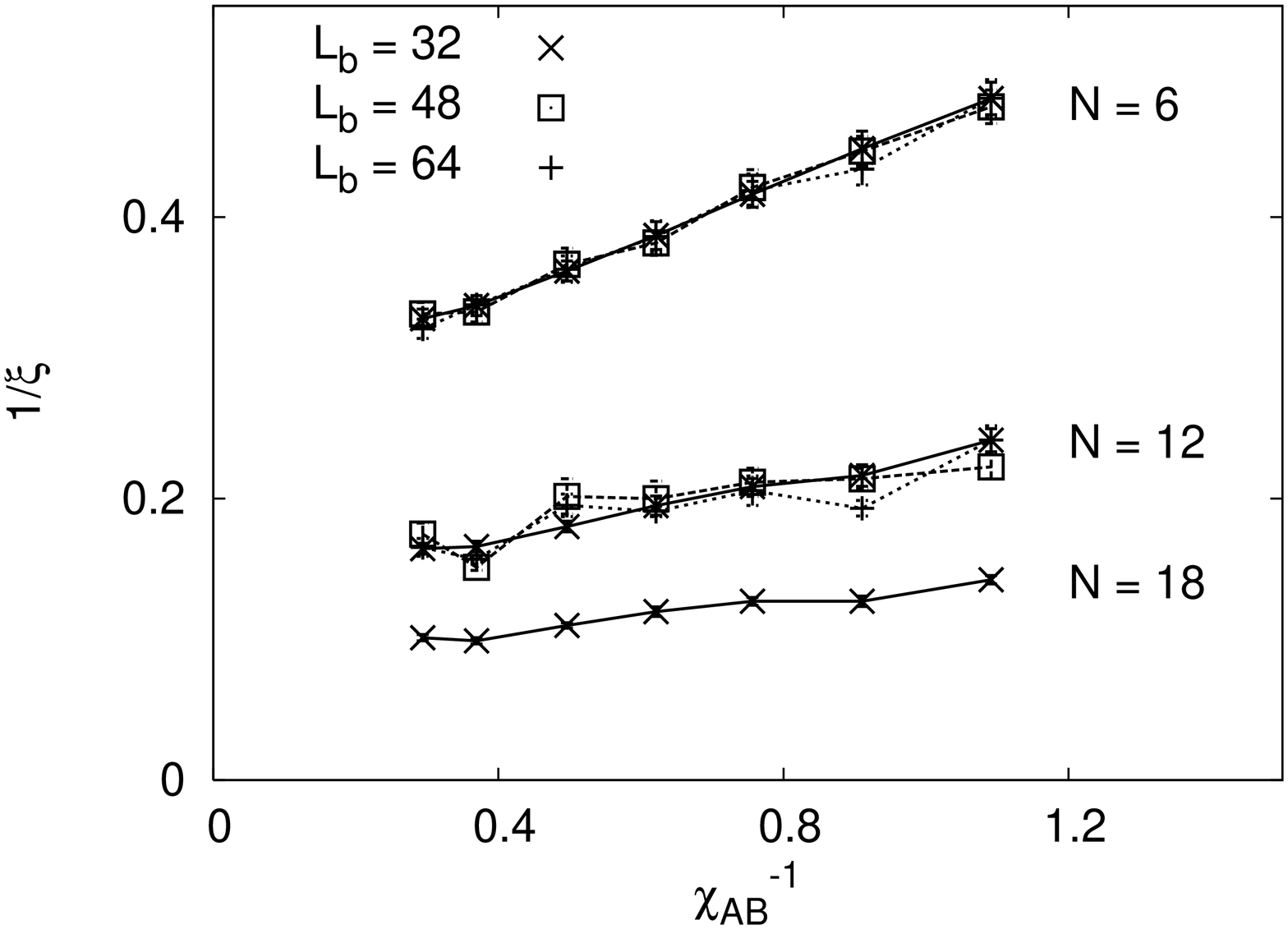, width=4.5cm, angle=270} \\
\end{array}$
\caption{Inverse of correlation lengths $\xi$ vs.\ $\chi_{AB}^{-1}$ for
copolymer bottle brushes with backbone length $L_b=32$, $48$, and $64$,
and three chain lengths $N=6$, $12$, and $18$.
Both good (left) and poor (right) solvent conditions are shown.}
\label{xidata}
\end{center}
\end{figure*}
                                                                                         
\begin{figure*}
\begin{center}
$\begin{array}{c@{\hspace{0.5cm}}c}
\epsfig{file=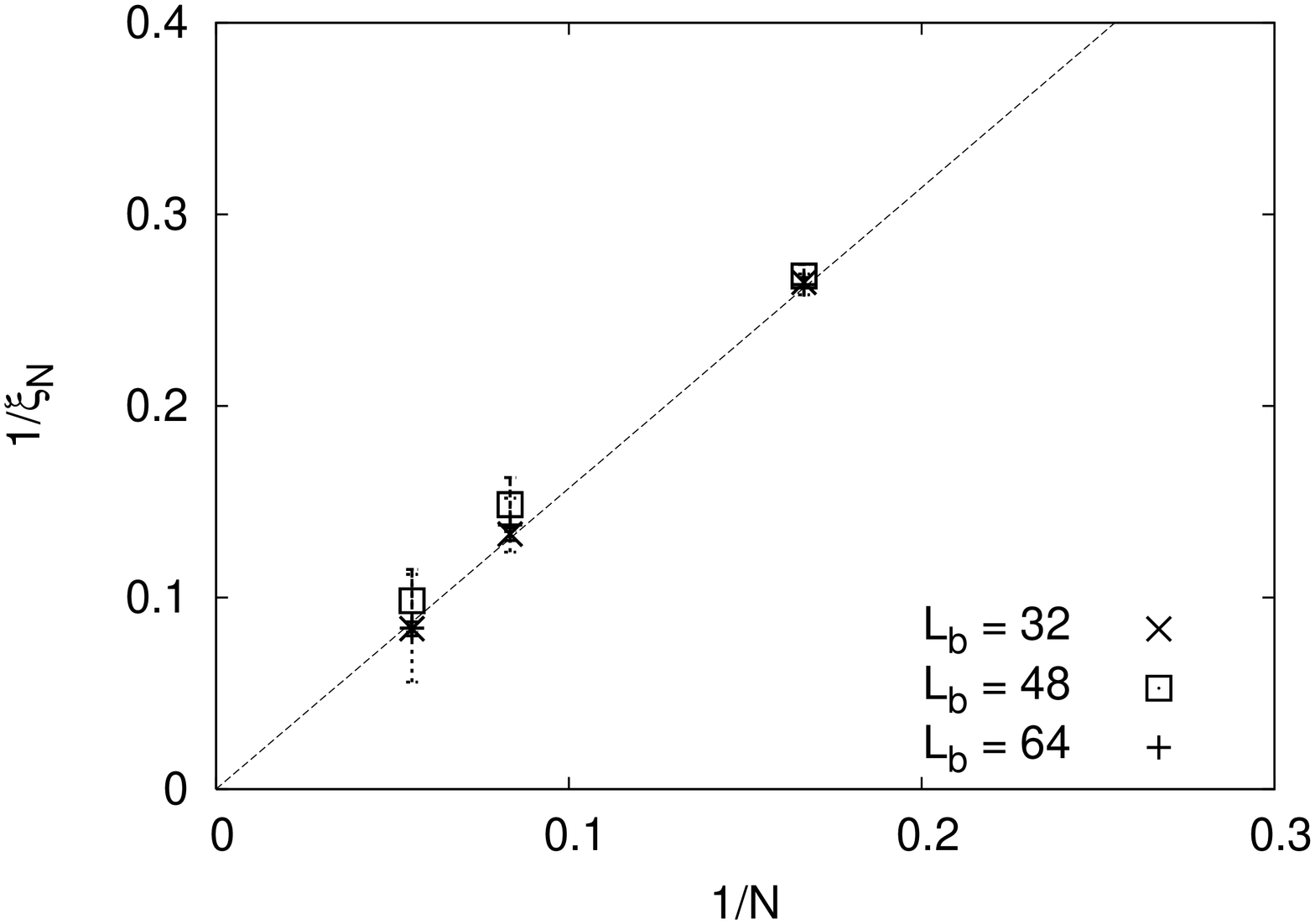, width=4.5cm, angle=270} &
\epsfig{file=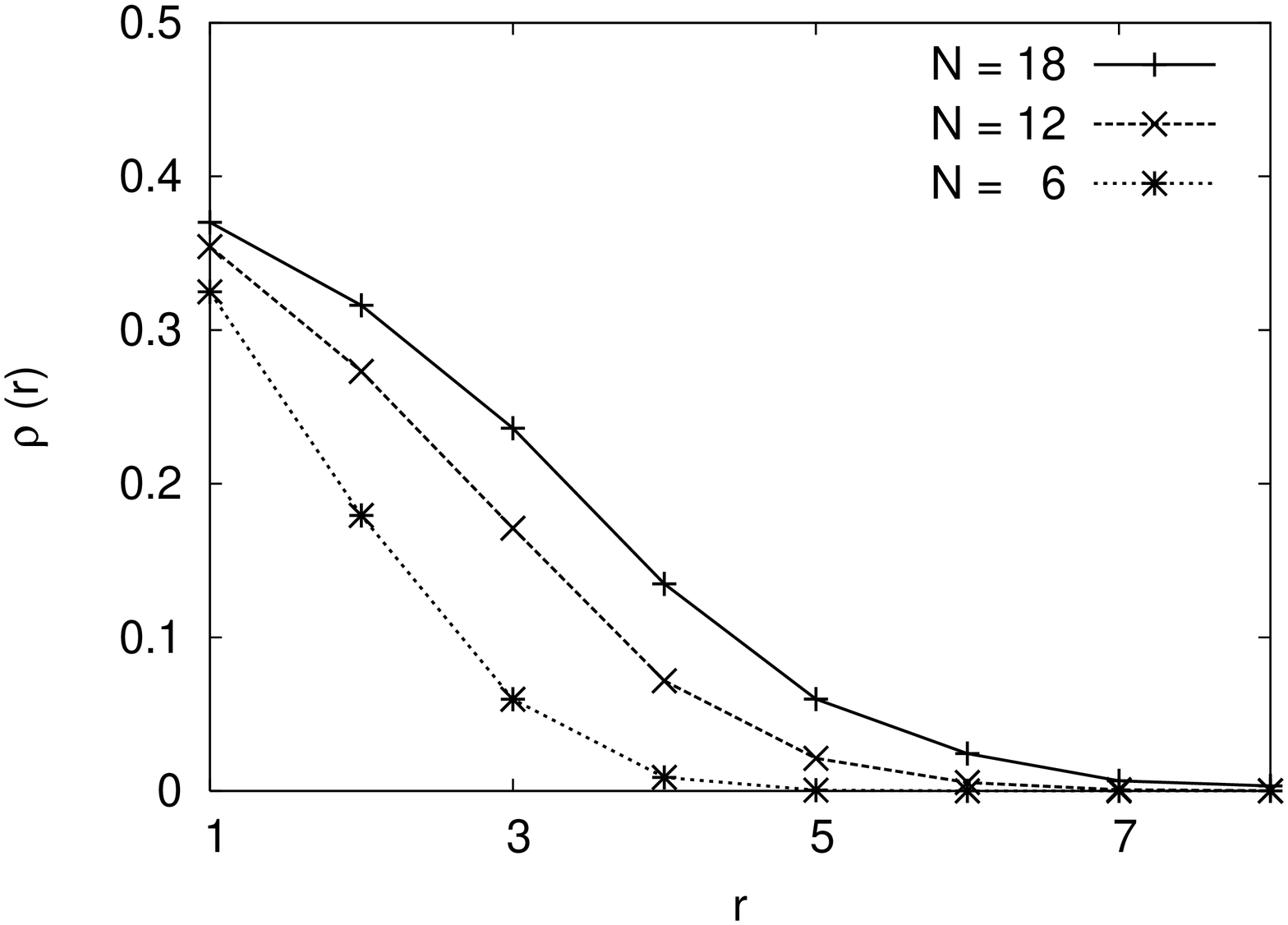, width=4.5cm, angle=270} \\
\end{array}$
\caption{Inverse of correlation length $\xi_N$ in the limit of
$\chi_{AB}\rightarrow \infty$ vs.\ $1/N$ for a poor solvent ($q=1.5$).
The straight line shows the asymptotic behavior as $N\rightarrow \infty$
(left part).
Monomer density profile $\rho(r)$ plotted vs.\ $r$ for a poor solvent ($q=1.5$),
$q_{AB}=0.1$ and $L_b=64$ (right part).} 
\label{rho}
\end{center}
\end{figure*}

\section{Results and Discussions}

As has also been discussed in~\cite{deJong}, the quantification of the phase
separation that occurs in a bottle brush is difficult. We follow
here an approach that is motivated by the analogy with the spin
problem mentioned in the introduction (fig.~\ref{cylinder}). 
We draw from each grafting site the vector $\vec{R}_{\rm cm}^{\alpha}$
to the center of mass (CM) of each side chain of type $\alpha$
($\alpha=A$ or $B$), and use the $x$ and $y$-components of these
vectors to define a unit vector $\vec{S}_i^{\alpha}$ that points
along that direction in the xy-plane. Making use of the fact
that we have grafted $A$ and $B$ chains in an alternating way, we compute
the correlation functions $<{\vec S}_i^A \cdot {\vec S}_{i+n}^A>$,
$<{\vec S}_{i+1}^B \cdot {\vec S}_{i+1+n}^B>$, for 
$n=0,\;2,\;4,\;6,\;\ldots \;$.
If we would have a symmetry breaking of "Janus cylinder" type
(i.e. a plane containing the backbone exists, such that the 
$A$-monomers are on one side of this plane and the $B$-monomers
on the other), the above correlations would not decay to zero as 
$n$ increases, but rather settle down at a nonzero order parameter square.
Figs.~\ref{cn-good} and ~\ref{cn-poor} show that this is not the case:
for good solvent conditions
$q=1$ even strong incompatibility ($q_{AB}\rightarrow 0$)
can induce only a correlation of rather short range.
In poor solvents there is also little
correlation when the side chains are very short ($N=6$) and
much stronger correlations occur for longer chains ($N=18$).
Although our data are already somewhat affected
by the finite length of the backbone (note the symmetry of
$C_n=C_{L_b-n}$), they are fitted by $\exp(-n/\xi)+\exp(-(L_b-n)/\xi)$ 
quite well.  
In fig.~\ref{snapshot} corresponding snapshots (similar to those of~\cite{deJong}) 
show rather pronounced phase separation, but this clearly is an
artifact of the finite system size. 
The correlation lengths $\xi$ which are determined by
fitting $C_n$ for 
$n\geq 2$ as shown in fig.~\ref{cn-good}
and ~\ref{cn-poor} are 
shown in fig.~\ref{xidata}.  As expected in view
of the discussion given in the introduction, $\xi$ increases
gradually with increasing chemical incompatibility $\chi_{AB}$
(decreasing temperature). 
Choosing poor solvent and/or $N$ larger makes $\xi$ also larger. 
Since $1/\xi$ decreases linearly as $\chi_{AB}^{-1}\rightarrow 0$,
one can extrapolate the data to $\chi_{AB}^{-1}=0$ by fitting
a straight line $1/\xi=1/\xi_N+b/\chi_{AB}$. 
Results of $\xi_N$ in a poor solvent are shown in fig.~\ref{rho}. 
It gives an indication that a sharp phase transition 
($1/\xi_N \rightarrow 0$) will develop
in the limit of large $\chi_{AB}$ and large $N$.
For very poor solvent condition,
one expects that the brush fills a cylinder of radius
proportional to $\sqrt{N}$ densely, so the situation should be
analogous to Eq.~(\ref{xispin}),
with $\pi R^2$ being replaced by $N$.
Our data are roughly compatible with such an interpretation.
However, the radial density profiles of the bottle brush
molecules (fig.~\ref{rho}) show that the actual density
profile $\rho(r)$ is still far from a profile of a densely
filled cylinder ($\rho(r)=1$ for $0<r<R$). This low density
is probably responsible for the fact that $\xi$ stays
finite as $\chi_{AB}^{-1}\rightarrow 0$ (fig.~\ref{xidata}).

   Alternatively, one could partition a bottle brush along its
backbone direction into disks with height $a$ (distance between
the neighboring grafting sites in our model). Each disk
only contains one grafting site.
The unit vector $\vec{S}_i^\alpha$ is then defined by
the direction from each grafting site to the CM of monomers of
type $\alpha$ ($\alpha=A$ or $B$) in the x-y plane.    
However, our results show that the correlation functions of
this observable decay much
faster due to the large fluctuations of the orientation of the vector
$\vec{S}_i^\alpha$.  

   Summarizing these results, we have presented evidence that the phase
separation towards a "Janus cylinder" structure develops completely
gradually as the incompatibility between the two types of monomers
increases. For our short side chain lengths, no trace of the sharp
phase transitions predicted by Stepanyan et al.~\cite{Stepanyan} could be
detected (a similar conclusion was also reached in~\cite{deJong}). Very large
length of side chains and/or very strong incompatibility is needed
to produce a correlation extending over a large number of side chains.
The fluctuation that has been ignored by the mean field theory~\cite{Stepanyan},
long wave length random "twist"-like rotation of the local interface between
the $A$-rich and $B$-rich region in the "Janus cylinder", costs very 
little energy and destroys long range order dramatically (fig.~\ref{cylinder}b).
On the other hand, varying suitably side chain length and solvent
quality one can control the nanoscopic length scale over which
Janus-type phase separation occurs along the backbone of the bottle
brush. For asymmetric side chain lengths (and flexible backbones)
this length scale is related to the spontaneous curvature of the 
cylinder-shaped bottle brush polymer. Controlling this curvature
is desirable for the anticipated application of such molecules in 
nanotechnology~\cite{Stephan}.

\section{Acknowledgments}

H.-P. H. thanks Prof. Peter Grassberger and Dr. Walter Nadler for very 
useful discussions. K. B. thanks M. Schmidt for stimulating discussions.
This work was financially supported by the Deutsche 
Forschungsgemeinschaft (DFG), SFB 625/A3.

\end{document}